\documentstyle[aps,preprint]{revtex}
\begin{document}
\setcounter{page}{1}
\setcounter{equation}{0}
\vskip 1truecm

\title{Correlations in Ising chains with non-integrable interactions}

\vspace {1.5 truecm}
\author{Birger Bergersen$^*$, Zolt\'an R\'acz$^\dagger$ and
Huang-Jian Xu$^\ddagger$}

\vspace{0.5 in}

\address{
$^*$Department of Physics, University of British Columbia,
Vancouver BC V6T 1Z1, Canada}

\address{
$^\dagger$Institute for Theoretical Physics, E\"otv\"os
University, 1088 Budapest, Puskin u. 5-7, Hungary}

\address{
$^\ddagger$Department of Physics, University of Manitoba, Winnipeg MB,
R3T 2B2, Canada}

\maketitle
\vspace{1in}
\begin{abstract}

Two-spin correlations generated by interactions which
decay with distance $r$ as $r^{-1-\sigma}$ with $-1 <\sigma <0$
are calculated for periodic Ising chains of length $L$.
Mean-field theory indicates that the correlations, $C(r,L)$, diminish
in the thermodynamic limit $L\rightarrow \infty$, but they
contain a singular structure for $r/L\rightarrow 0$  which can be
observed by introducing magnified correlations, $LC(r,L)-\sum_rC(r,L)$.
The magnified correlations are shown to have a
scaling form $\Phi (r/L)$ and the singular structure of $\Phi (x)$
for $x\rightarrow 0$ is found to be the same at all temperatures including
the critical point. These conclusions are supported by the results
of Monte Carlo simulations for systems with $\sigma =-0.50$ and $-0.25$
both at the critical temperature $T=T_c$ and at $T=2T_c$.

\vspace{1cm}

\end{abstract}
\pagebreak
\section{Introduction}

The Ising model in $d=1$ dimension with long-range
interactions decaying with distance $r$ as $J(r)\sim r^{-1-\sigma}$
has been much studied. Dyson \cite{A157}
showed that there is no phase transition for $\sigma>1$, while there
is long-range order at low temperatures
when the interactions are ferromagnetic and
$\sigma\leq 1$. The borderline case of $\sigma=1$ was studied
by several authors \cite{676}, who found a
one-dimensional version of the Kosterlitz-Thouless transition
\cite{347} in the system.
A variety of methods \cite{{672},{690},{639},{670}}
have been used to show that the transition is of second-order
with continuously varying critical exponents in the region
$\frac{1}{2}<\sigma<1$. For
$0<\sigma<\frac{1}{2}$, it has been found that the critical exponents
are mean field like for thermodynamic quantities, but the spin-spin
correlations at the critical temperature
\begin{equation}
C(r)\propto\frac{1}{r^{d-2+\eta}}=\frac{1}{r^{\eta -1}}
\end{equation}
have a non-classical critical exponent $\eta=2-\sigma$
\cite{{675},{673}}. The case $\sigma=\frac{1}{2}$ is somewhat special
in that the mean field exponents are modified by logarithmic
corrections \cite{672} in analogy to the situation with the short
range Ising model for $d=4$, or the $d=3$ Ising model with dipolar
interactions \cite{667}.

The case $\sigma<0$ does not appear to have been investigated except
for $\sigma =-1$. The reason, perhaps, is the fact that the
interaction potential, $J(r)$,
becomes nonintegrable for $\sigma<0$ and then the energy of the system
is non-extensive.
The textbook example of $\sigma=-1$ has attracted much attention
since it corresponds to an Ising model in which every spin
interacts equally with every other spin. The problem of
non-extensive energy is solved by rescaling  the coupling strength
by the number of spins and the result is a mean-field exact solution.
Spatial correlations have no meaning in this system.

For $\sigma <0$ and $\sigma\neq -1$
one can still consider spatial correlations and this is what we
shall do in this paper. The reason for doing it is partly
the fact that there are interactions in nature which are non-integrable.
The gravitational interaction is a prominent example but there are
many systems which contain topological defects with non-integrable
effective interactions between the defects.
Our original motivation, however, comes from the studies of
nonequilibrium steady
states. The main difficulty with nonequilibrium steady states
is that their properties depend not only on the interactions but also
on the details of the underlying dynamics. In order to get a handle on the
effect of dynamics, there has been several works
\cite{{DRT},{BirR},{Schm},{BassR}} in which the goal was to
determine the effective interactions which are generated by the interplay
of various dynamical processes. A way to achieve this goal is to investigate
the phase transitions occurring in nonequilibrium steady states and
deduce the effective interactions from the universality class the
transitions fall into. For example, a combination of spin-flips and
Levy-flight spin exchanges in a  $d=1$ kinetic Ising model
produces a steady state with effective long-range interactions and
this steady state displays a critical phase transition. In this case,
the large-distance behavior of the interactions
$J(r)\sim r^{-1-\sigma}$ can be deduced \cite{BirR}
from measuring critical exponents. Of course, this can be done only if
$\sigma$ is in a range $0<\sigma <1$  where either the
thermodynamic exponents or the correlation exponent $\eta$ depend
on $\sigma$. In order to extend this method to
non-integrable interactions
$(\sigma <0)$, we must investigate the correlations since measuring
thermodynamic exponents yield mean-field values independently of $\sigma$.
Thus, we can see that in order to extract information about non-integrable
effective interactions in nonequilibrium steady states, we must first
understand the {\em equilibrium} correlations in systems with non-integrable
interactions.

Since mean-field theory is expected to
describe the critical behavior correctly for $\sigma <0$,
we used it to calculate the correlations for finite size systems
(section \ref{meanf}). Our main result is that the singular part of
the correlations appears to have universal features which make it useful
in deducing effective interactions in nonequilibrium steady states.
We check our mean-field results against simulations in section \ref{montec}
and a final discussion is given in section \ref{final}.

\section{Calculation of the correlation function in mean-field theory}
\label{meanf}
We consider a one-dimensional Ising model of $L=2N+1$ spins $s_i =\pm 1$
at sites $i=-N,-N+1,..., N$ and impose periodic boundary conditions.
The energy of the system is given by
\begin{equation}
{\cal H}=-\sum_{i<j}J_{i,j}s_is_j-\sum_iH_is_i \quad ,
\end{equation}
and the interaction is assumed to decay with distance as a power law
but it is cut off at $\vert i-j \vert=N$:
\begin{equation}
J_{i,j}=\left\{\begin{array}{ccc}
J/\vert i-j \vert^{1+\sigma}&\;{\rm for}\;&1\le\vert i-j\vert\le N \\
0& &{\rm otherwise.}\\
\end{array}\right.
\end{equation}
Here $J>0$ and the parameter $\sigma$ is assumed to be in the range
$-1<\sigma <0$. The inhomogeneous external field, $H_i$
is included in order to calculate the susceptibility
$\chi_{i,j}=\partial \langle s_i \rangle / \partial H_j$ which,
at temperature $\beta^{-1}=k_BT$, is related to the two-spin correlation
function $C_{i,j}=\langle s_i s_j \rangle $ through
\begin{equation}
C_{i,j}=k_BT\left.\frac{\partial \langle s_i \rangle}{\partial H_j}
\right|_{\{ H_l \} \rightarrow 0} \quad ,
\label{fluctdiss}
\end{equation}
where ${\{ H_l \} \rightarrow 0}$ means that the susceptibility and,
consequently, the correlation functions are evaluated in the limit of
vanishing fields.

In the mean-field approximation, $\chi_{i,j}$ is obtained from the
self-consistency equation for the average magnetization:
\begin{equation}
\langle s_i \rangle =\tanh \left (
\sum_l \beta J_{i,l}\langle s_l \rangle +
\beta H_i \right )\quad .
\label{tanh}
\end{equation}
Taking a derivative of both sides of this equation with respect to $H_j$
and letting ${\{ H_l \} \rightarrow 0}$, we obtain
\begin{equation}
C_{i,j}= \left [\sum_l \beta J_{i,l}C_{l,j}+ \delta_{i,j} \right ]
(1-\langle s_i \rangle^2) \quad .
\label{Cij}
\end{equation}
Equations (\ref{tanh}) and (\ref{Cij}) from a closed set of equations
for $C_{i,j}$ and $\langle s_i \rangle$.
In zero field, the system is expected to be homogeneous
and thus $C_{i,j}=C_{i-j}$ and $\langle s_i \rangle =m$ where $m$ is given
by the stable solution of (\ref{tanh}).
Assuming this homogeneity, equation (\ref{Cij}) is solved
by Fourier transformation. Introducing
\begin{equation}
C(k)=\sum_{l=-N}^{N} C_l\exp(-ikl); \quad k=\frac{2\pi n}{2N+1}; \quad
n=0,\pm1,...,\pm N \quad ,
\label{C(k)}
\end{equation}
one finds
\begin{equation}
C(k)=\frac{1-m^2}{1-(1-m^2)\beta J(k)} \quad ,
\label{Cfourier}
\end{equation}
where J(k) is the Fourier transform of $J_{i,j}=J_{i-j}$.

Up to this point, the derivation followed along standard lines.
The non-integrability of the interaction starts to play role when
we try to find the critical temperature, $T_c$. Since $C(k=0)$ is proportional
to the fluctuation of magnetization, its divergence determines $T_c$ in the
thermodynamic limit $N\rightarrow \infty$. Thus setting $m=0$ in
eq.(\ref{Cfourier}) one finds
\begin{equation}
k_BT_c=J(0)=\lim_{N\rightarrow \infty}\sum_{n=1}^{N}\frac{2J}{n^{1+\sigma}}
 \quad .
\label{Tc}
\end{equation}
The problem with the above expression is that the sum diverges as
$N^{\vert \sigma \vert}$ for $\sigma <0$. Thus, in order to have a
thermodynamic limit for the correlation functions, we either have to
divide $J$ by
$N^{\vert \sigma \vert}$ or have to consider temperatures which are
proportional to $N^{\vert \sigma \vert}$. Since only the ratio $J/k_BT$
enters the expressions, the two routes are equivalent. Introducing
$\theta =T_c(N)/T$ and ${\bar\theta} =(1-m^2)\theta$, where $T_c(N)$ is
defined through
the sum in ($\ref{Tc}$) without taking the $N\rightarrow \infty$ limit,
we obtain the spatial correlation function from the inverse Fourier transform
of (\ref{Cfourier}):
\begin{equation}
C_l=\frac{1-m^2}{2N+1} \left \{ \frac{1}{1-{\bar\theta}}+
2\sum_{n=1}^N \frac{\cos{\frac{2\pi n}{2N+1}l}}{1-{\bar\theta}Q_N(n)} \right \}
\quad ,
\label{Cn}
\end{equation}
where  $Q_N(n)$ is given by
\begin{equation}
Q_N(n)=\sum_{j=1}^N \frac{ \cos{ \frac{2\pi j}{2N+1} n} }{j^{1+\sigma}}/
\sum_{j=1}^N \frac{1}{j^{1+\sigma}}
\quad .
\label{Qn}
\end{equation}
One can see that correlations in the low- and high-temperature phases
are simply related by rescaling the temperature and the amplitude
by $1-m^2$ \cite{foot1}. For this reason we shall consider only the
high-temperature phase and the critical point and thus set $m=0$ and
$\bar\theta =\theta$.

Equations (\ref{Cn}) and (\ref{Qn}) can be easily evaluated numerically and
the mean-field results can be compared to MC simulations (see next Section).
First, however, we shall discuss the analytic properties of $C_l$ in the
``thermodynamic limit" where $N\rightarrow \infty$.  As can be seen
from (\ref{Cn}) the average of $C_l$ over $l$ is given by
$[(2N+1)(1-\theta )]^{-1}$ and is negligible
for $\theta\not= 1$. The part of $C_l$ which remains after subtracting
the average also diminishes but with a smaller exponent. It is of the
order of $N^\sigma$ and thus no spatial correlations remain in the
$N\rightarrow \infty$ limit for $\sigma <0$. It turns out, however, that
the correlation function has an interesting singular behavior for small
$x=l/(2N+1)$ which can be well observed in finite-size systems.
We shall investigate this spatial dependence
by amplifying it through subtracting the average part of $C_l$ and
multiplying the result by $2N+1$:
\begin{equation}
\Phi_N(x)\equiv (2N+1)C_l- \frac{1}{1-\theta} =
2\sum_{n=1}^N \frac{\cos{(2\pi nx)}}{1-\theta Q_N(n)}
\quad .
\label{Phi}
\end{equation}
The function $Q_N(n)$ is finite and can be evaluated exactly
in the $N\rightarrow \infty$ limit
\begin{equation}
\lim_{N\rightarrow \infty}Q_N(n)= \frac{\pi^\sigma
\Gamma{(1- \sigma )} \cos{\frac{\sigma \pi }{2}} }
{ n^{\vert \sigma \vert}} \equiv \frac{D_\sigma }{n^{\vert \sigma \vert}}
\quad ,
\label{Qninf}
\end{equation}
where $\Gamma (z)$ denotes the gamma function. Thus
we can take the upper limit of the sum in (\ref{Phi}) to infinity and
find a scaling function, $\Phi(x)$, which is independent of $N$. The
singular part of $\Phi(x)$ in the limit
$x\rightarrow 0$ is calculated by first noting that
the identity $2\sum_{n=1}^N\cos{(2\pi nx)} =-1$
is valid for any $x$ that can be written as $l/(2N+1)$ where $l$
is an integer not divisible by $2N+1$. Using this identity,
the sum in (\ref{Phi}) can be rewritten so that it contains the expression
$(1-\theta D_\sigma /n^{\vert \sigma \vert})^{-1}-1$. This expression
is then expanded in $\theta$ and, finally,
the sums over $n$ are changed into integrals.
The number of singular terms obtained in this way depends on
$\sigma$ and, for $(j+1)^{-1}<\vert
\sigma \vert <j^{-1}$, where $j$ is an integer, we find:
\begin{equation}
\Phi_{sing}(x)=\theta \frac{A_1(\sigma )}{x^{1+\sigma}}+
\theta^2 \frac{A_2(\sigma )}{x^{1+2\sigma}}+\ldots +
\theta^j \frac{A_j(\sigma )}{x^{1+j\sigma}}
\quad .
\label{Phising}
\end{equation}
The coefficients $A_n(\sigma )$ can be calculated explicitly. For example,
the coefficient of the strongest singularity, $A_1$, is given by
\begin{equation}
A_1(\sigma )=2^{-1- \sigma } \vert \sigma \vert
\quad .
\label{A1}
\end{equation}
If $\vert \sigma \vert =1/j$ then the last term in (\ref{Phising})
changes from power-law
singularity into a term of logarithmic singularity, $A_j\ln{(x)}$. For example,
the case $\sigma =-1/2$ yields the following singular part
\begin{equation}
\Phi_{sing}(x)=\frac{\theta }{\sqrt{8x}}+
\frac{\theta^2}{4}\ln{(x)}
\quad .
\label{Phi1/2}
\end{equation}
One can compare $\Phi_N(x)$ calculated for $\sigma =-0.5$ and $\theta =0.5$
from the exact sum with e.g. $2N+1=801$
and $\Phi_{sing}(x)$ obtained from (\ref{Phi1/2}). One then finds
that $\Phi_N(x)-\Phi_{sing}(x)$ is constant $(A=-2.35)$ to
an accuracy of the order of $1\%$.
Thus the singular part plus a constant represents
the exact solution to a very good accuracy over
the entire range, $0<x<0.5$, which is relevant physically.
We checked the ``singular-terms-plus-constant"
description for other values of $\sigma$ and found that the accuracy
decreases as $\sigma $ decreases.
For example, in order to achieve a $1\%$ accuracy
at $\sigma =-0.75$ and $\theta =1$, one must also include an additional term
and write $\Phi (x)\approx A_1x^{-1/4}(1+Bx) +C$.

Next, we turn to the calculation of $\Phi(x)$ at the critical point
($\theta =1$).
In systems with short-range interactions, the decay of correlations
changes from exponential to power law as the critical point is reached.
Systems with long-range, integrable interactions ($\sigma >0$) have
power-law correlations
already in the high temperature phase and the change that occurs in these
systems when $T_c$ is approached is that the exponent of the
power law decay changes (becomes smaller at $T_c$). For example,
in the exactly solvable
spherical model \cite{Joyce}, $C_l\sim l^{-d-\sigma}$ for $T>T_c$ and $C_l\sim
l^{-d+\sigma}$ at $T_c$. As we shall see below, the non-integrable systems are
different from the cases discussed above in the sense that the singular
structure of $C_l$ does not change when $\theta =1$ is reached. The scaling
function $\Phi_{sing} (x)$ at criticality is obtained by just setting
$\theta =1$ in eq.(\ref{Phising}).

The first question that arises when the $\theta =1$ case is considered is the
infinity of the $1/(1-\theta )$ term which has been subtracted from
$(2N+1)C_l$ in the definition of $\Phi$ (see (\ref{Phi})).
Since this term is the sum of the correlation function, it is equal
to the magnetization fluctuation:
\begin{equation}
\frac{1}{1-\theta}=C(k=0)=\frac{<M^2>}{2N+1}
\quad ,
\label{mfluct}
\end{equation}
where $M=\sum_is_i$ is the total magnetization of the
system. Thus, at the critical point, $(1-\theta )^{-1}$ does not diverge in a
finite-size system,
it just changes from a quantity of $O(1)$ to $O(N^\rho)$ where $\rho =1/2$ in
the mean-field theory. The apparent singularity of $1/(1-\theta )$ is the
result of effectively linearizing eq.(\ref{tanh}) when using
eq.(\ref{fluctdiss})
for calculating the correlation function. For our purpose of calculating
$\Phi$,
the real or apparent singularity of $C(k=0)$ does not pose a difficulty
since $\Phi$ is defined after subtracting $C(k=0)$ from the correlation
function. Note that $\Phi (x)$ is well defined physically: it is the
correlation function after the average has been subtracted.

The second question that has to be answered for $\theta =1$ is whether
any of the terms in the sum (\ref{Phi}) defining $\Phi$ becomes singular
in the $\theta \rightarrow 1$ limit (After all $1/(1-\theta )$ was a
term in this sum before being treated separately and, in general,
one may expect long-wavelength singularities at the critical point).
The answer to this question is {\em no} and the reason for the absence
of singularities is that long-range interactions
develop a gap in the ``fluctuation spectrum" of the system.
In order to see how this happen and why non-integrable interactions
are different from the short-range ones, let us examine
the denominator of the terms in the sum in eq.(\ref{Cn}). For short-range, e.g.
nearest-neighbor, interactions one finds in the long-wavelength limit
\begin{equation}
1-\theta Q_N(n)=1-\theta \cos{ \frac{2\pi}{2N+1} n} \approx
1-\theta +\frac{\theta}{2}\left ( \frac{2\pi n}{2N+1} \right )^2
\quad .
\label{denom}
\end{equation}
One can see that the above expression is nonzero for $\theta <1$ while
it goes to zero at $\theta =1$ as the square of the wavenumber
(for $N\rightarrow \infty$).
This is the origin of the exponential (power law) decay of {\em large-distance}
correlations at $\theta < 1$ ($\theta =1$).

The structure of the denominator changes for non-integrable potentials
($\sigma <0$). It becomes independent of $N$  for $N\rightarrow \infty$
and, using (\ref{Qninf}), one finds
\begin{equation}
1-\theta Q_N(n) \approx
1-\theta\frac{\pi^\sigma
\Gamma{(1- \sigma )} \cos{\frac{\sigma \pi }{2}} }
{ n^{\vert \sigma \vert}}
\quad .
\label{denom2}
\end{equation}
since $D_\sigma\equiv \pi^\sigma
\Gamma{(1- \sigma )} \cos{\frac{\sigma \pi }{2}}<1$ for $-1<\sigma <0$
($D_\sigma\rightarrow 1$ only for $\sigma \rightarrow 0$), the denominator in
(\ref{denom2}) is nonzero even for the smallest $n=1$ at $\theta \leq 1$. Thus,
there is a gap in the spectrum of magnetization fluctuations for all
wavenumbers, and the singularity of the scaling function $\Phi (x)$ at
{\em small} arguments (small distances) comes from the non-analytic
structure of the denominator at large wave-numbers. It should be noted that
the above argument also holds in the low-temperature phase after replacing
$\theta$ by $\bar\theta$ and using the fact that $0\leq {\bar\theta}\leq 1$
\cite{foot1}.

It is clear from the above discussion that $\theta =1$ can be set in the sum
(\ref{Phi}) for evaluating $\Phi (x)$, and it is also clear that the result
for the singular part $\Phi_{sing}$ is given by eq.(\ref{Phising})
with $\theta =1$. One can again compare $\Phi_{sing}$ and $\Phi$, the latter
evaluated for large but finite systems and one again finds $1\%$ accuracy with
the $\Phi_{sing}+const$ description at $\sigma =-0.5$. The trends with
the accuracy of the singular-part description is also the same for $\theta =1$
as in the case of $\theta <1$ discussed above.

\section{Monte Carlo simulations}
\label{montec}
In this section, we describe the results of addressing the following question:
Do the results of the mean-field calculation
remain valid when fluctuations
are included? Since no exact solutions are available, we studied the problem
by performing MC simulations on the long-range Ising model
described at the beginning of Section II. We considered cases of
$\sigma =-0.50$ and $-0.25$ at both $\theta =0.5$ and $1.0$.
Single-spin-flip
dynamics with Metropolis flip rates was used and we studied systems of sizes
$2N+1=201,\, 401,\, 801$ and $1601$. After an initial estimate of the
relaxation times, we let the system equilibrate and the equilibrium
correlations were calculated and analyzed.

Apart from the usual difficulties of simulating a system with long-range
interactions (calculating the energy changes is time-consuming and no cluster
algorithms can be used to reduce critical slowing down), there is one extra
problem when a $\sigma <0$ system is simulated. Namely, the definition of
the $\theta =1$ critical point is not quite obvious. The mean-field
$T^{MF}_c(N)$
defined in (\ref{Tc}) has the right asymptotics for $N\rightarrow \infty$ but
for
a finite $N$, $T_c$ has corrections due to fluctuations. For a sequence of
systems of increasing size, one can, in principle, use
any sequence of $T_c(N)$ which is approaching $T^{MF}_c(N)$ in the
thermodynamic limit. In practice, we defined $T_c(N)$ as the temperature
where the macroscopic magnetization fluctuations have the large-$N$ behavior
that follows from mean-field theory, namely
$\langle M^2 \rangle /(2N+1) \approx 1.17..(2N+1)^{1/2}$. Since
$\langle M^2 \rangle /(2N+1) =1/(1-\theta )$, the above choice means that
we approach the $\theta_c =1$ critical point as $\theta_c(N)=
1-[1.17..(2N+1)]^{-1/2}$. The same correction can also be made in the
mean-field calculation. This correction, however, is small and would not
be visible in the Figures discussed below.

Fig.1 displays the raw data for the correlation function $C_l$
at $\sigma =-0.5$ and $\theta_c(N)$. After subtracting
the average correlation and multiplying the result by $2N+1$ (i.e.
calculating $\Phi$) we obtain a
collapse of data (Fig.2) if $\Phi$ is plotted against $x=l/(2N+1)$.
A blow-up of the small $x$ region is shown in Fig.3.

In Fig.4, the MC data is compared with
$\Phi (x)$ calculated from the mean-field theory of Sec.II, and we
observe excellent agreement between simulations and theory. On Fig.5, we
can see similar agreement away from the critical
point $(\theta =0.5)$ for the case of $\sigma =-0.5$.
We have also performed simulations with $\sigma =-0.25$ at both $\theta =0.5$
and $\theta_c(N)$ and we again observed both the scaling of the data when
$\Phi (x)$ vs. $x$ was plotted as well as the agreement between MC and
mean-field
results. Thus we conclude that the mean-field theory presented in Sec.II
correctly describes the correlations of Ising models in which the interactions
are of long-range, power law form $J_l\sim l^{-1-\sigma}$ with $-1<\sigma <0$.

\section{Final remarks}
\label{final}
The main conclusions of our work are that (i) the mean-field theory
provides us with explicit expressions for the correlation function in case
of nonintegrable interactions and (ii) the
singularity of the correlations for a given $\sigma$
has the same structure at all temperatures. The results appear to
be not affected by fluctuations as indicated by
Monte Carlo simulations for particular values of $\sigma$ and $T$.

We expect that the mean-field theory is also correct for higher
dimensional systems where the fluctuations play even less role.
This problem is currently being investigated.

\section*{ Acknowledgement}
Z.R. and B.B. are grateful for hospitality extended to them during
their stay at University of British Columbia and E\"otv\"os University,
respectively. This work was supported by Natural Science and
Engineering Research Council of Canada, by the
Hungarian Academy of Sciences, and by EC Network Grant N$^o$
ERB CHRX-CT92-0063.

\newpage

\centerline{\bf \Large List of figures:}
\vspace{1in}
\centerline{Figure 1:}

Raw data from the MC simulation of an Ising chain of length $2N+1$
at the critical temperature $T=T_c$ with the long-range parameter set
to $\sigma =-0.5$. The correlation function, $C_l$ is plotted against
$l$.

\centerline{Figure 2:}
Scaling of data in Fig.1 after subtracting the average of $C_l$ over $l$
and multiplying the results by $2N+1$, i.e. plotting
$\Phi_N(x)$ defined in eq.(\protect\ref{Phi}) against $x=l/(2N+1)$.
For clarity, only 40 points are included from each data sets.

\centerline{Figure 3:}
Blow-up of the small $x$ region of Fig.2. All data points in the given
$x$ interval are included.

\centerline{Figure 4:}
Comparison of the scaled MC data (Fig.2) and the mean-field
results for the $2N+1=1601$ data.

\centerline{Figure 5:}
Comparison of the scaled MC data and the mean-field
results for $\sigma =-0.5$ and $T=2T_c$.


\begin{thebibliography}{99}
\bibitem{foot1} It can be shown using (\ref{tanh}) that the rescaled
temperature
${\bar\theta}$ has the same range $0\leq {\bar\theta}\leq 1$ in both the
high- and low-temperature phases.
\end{thebibliography}
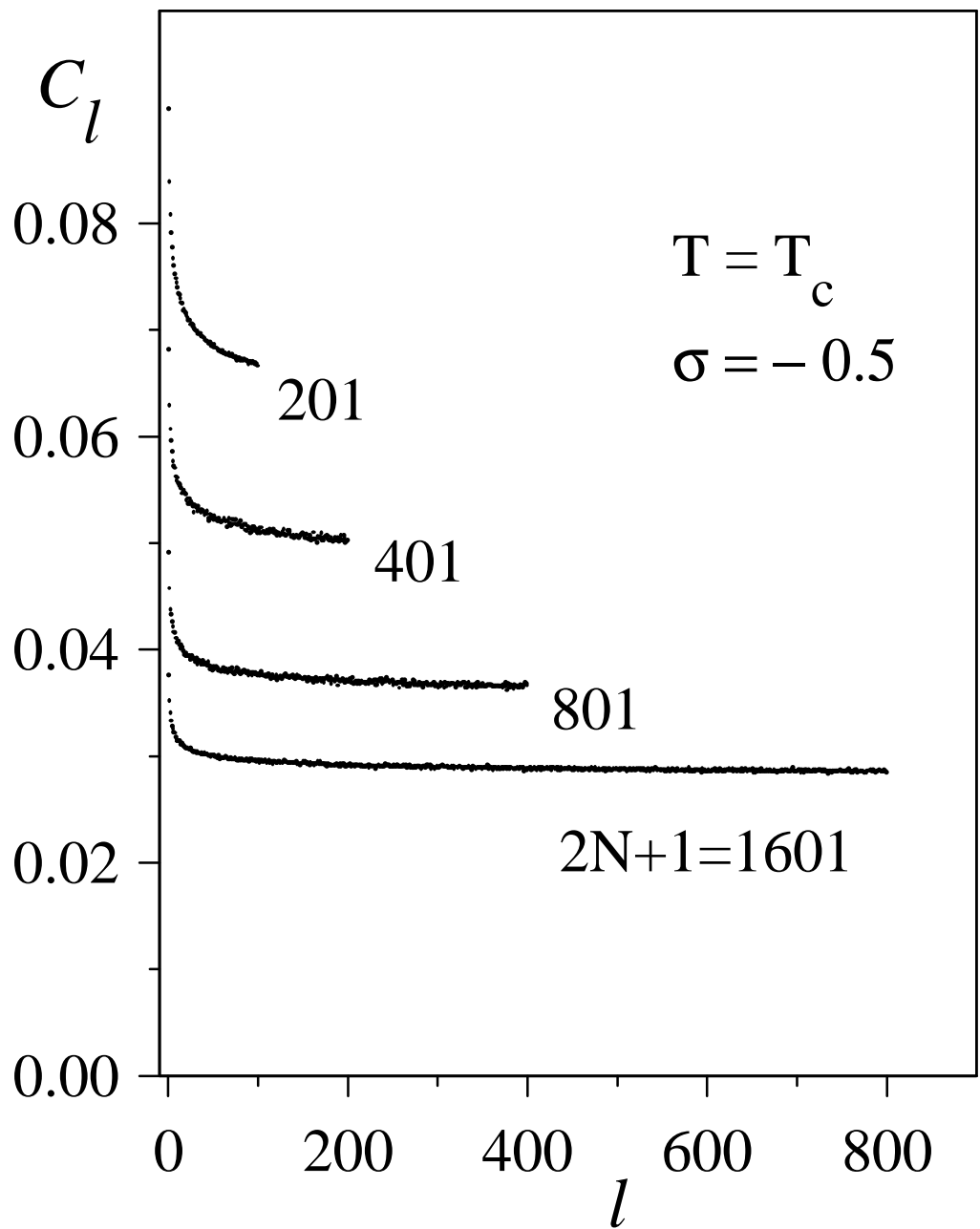
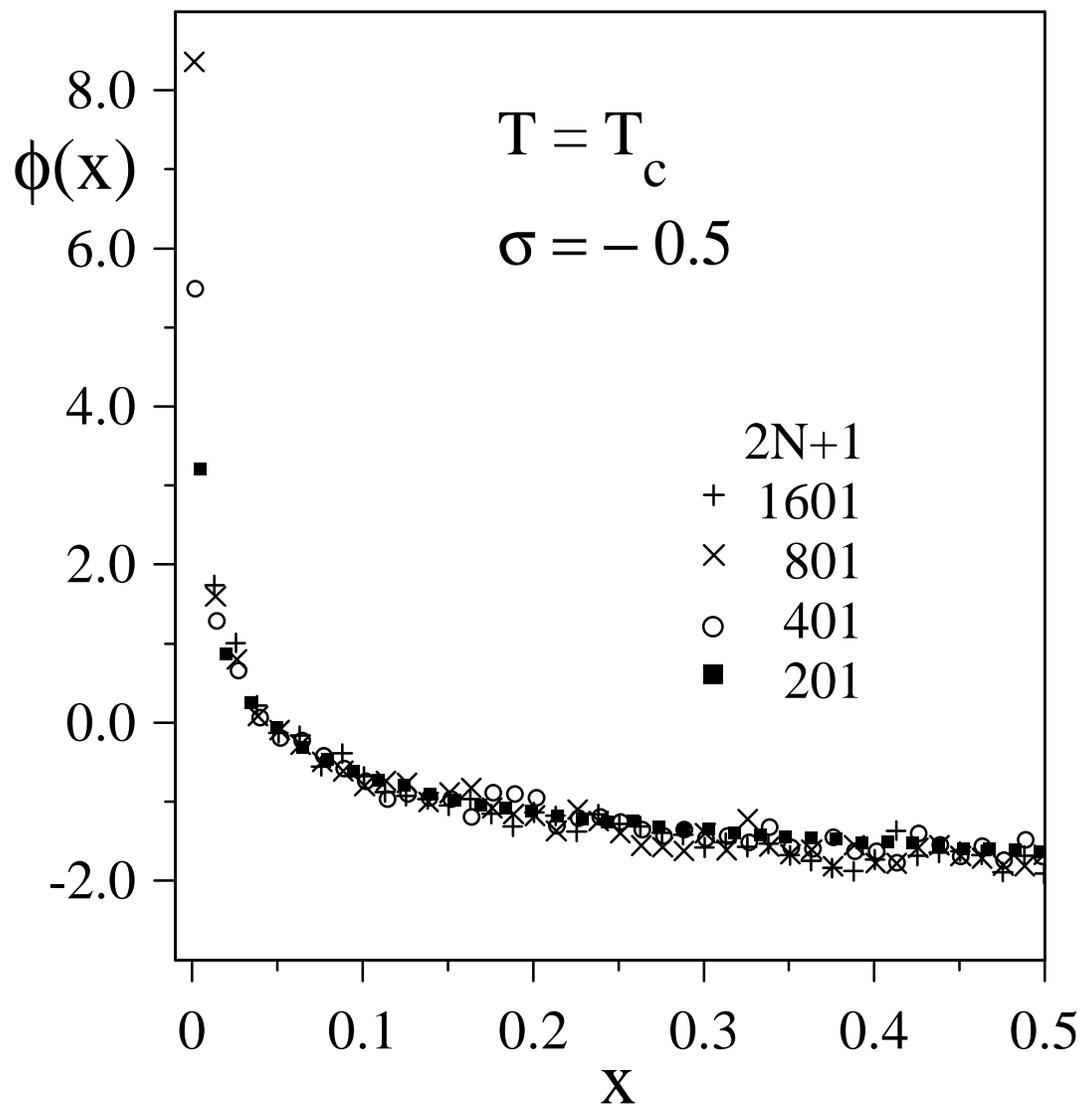
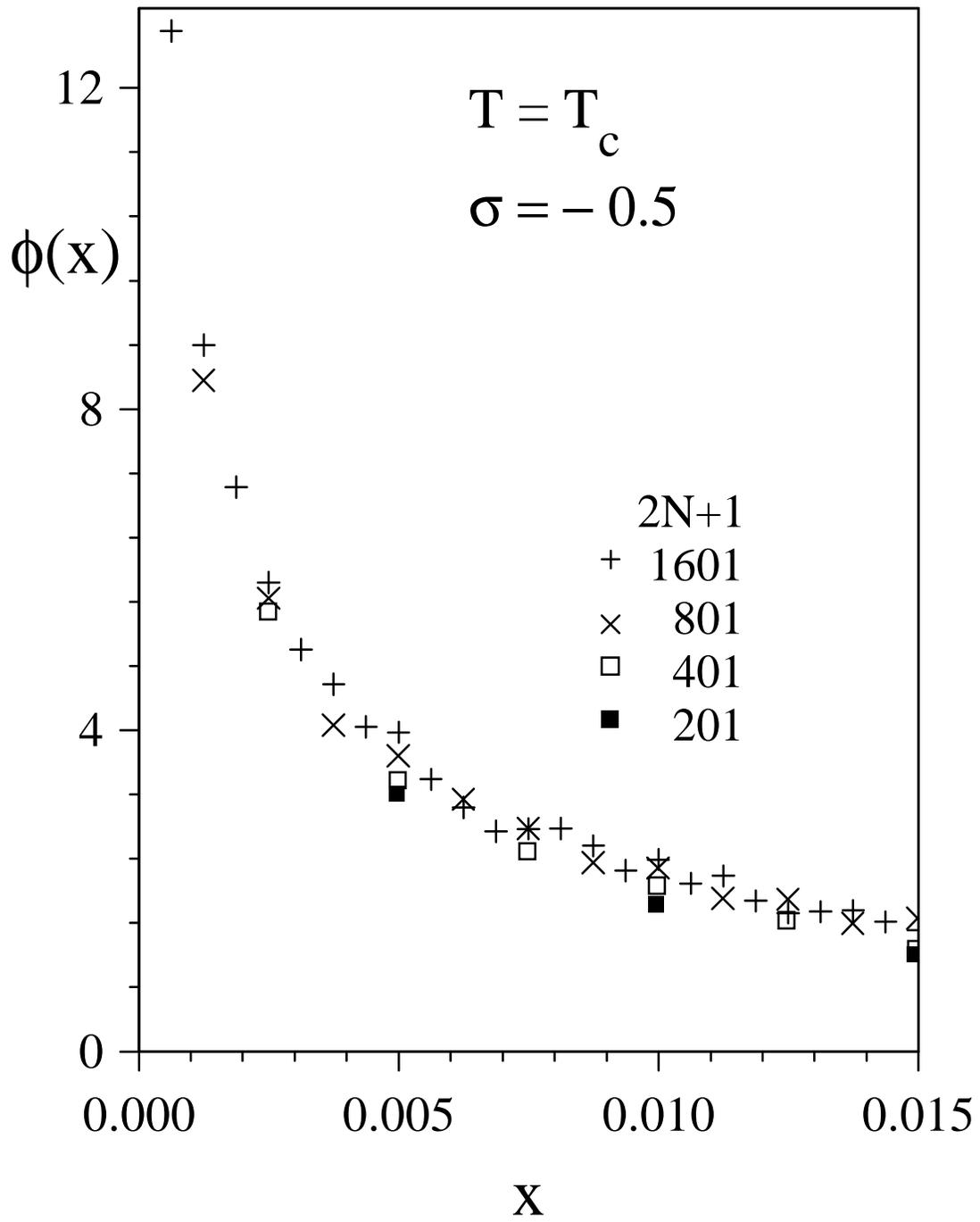
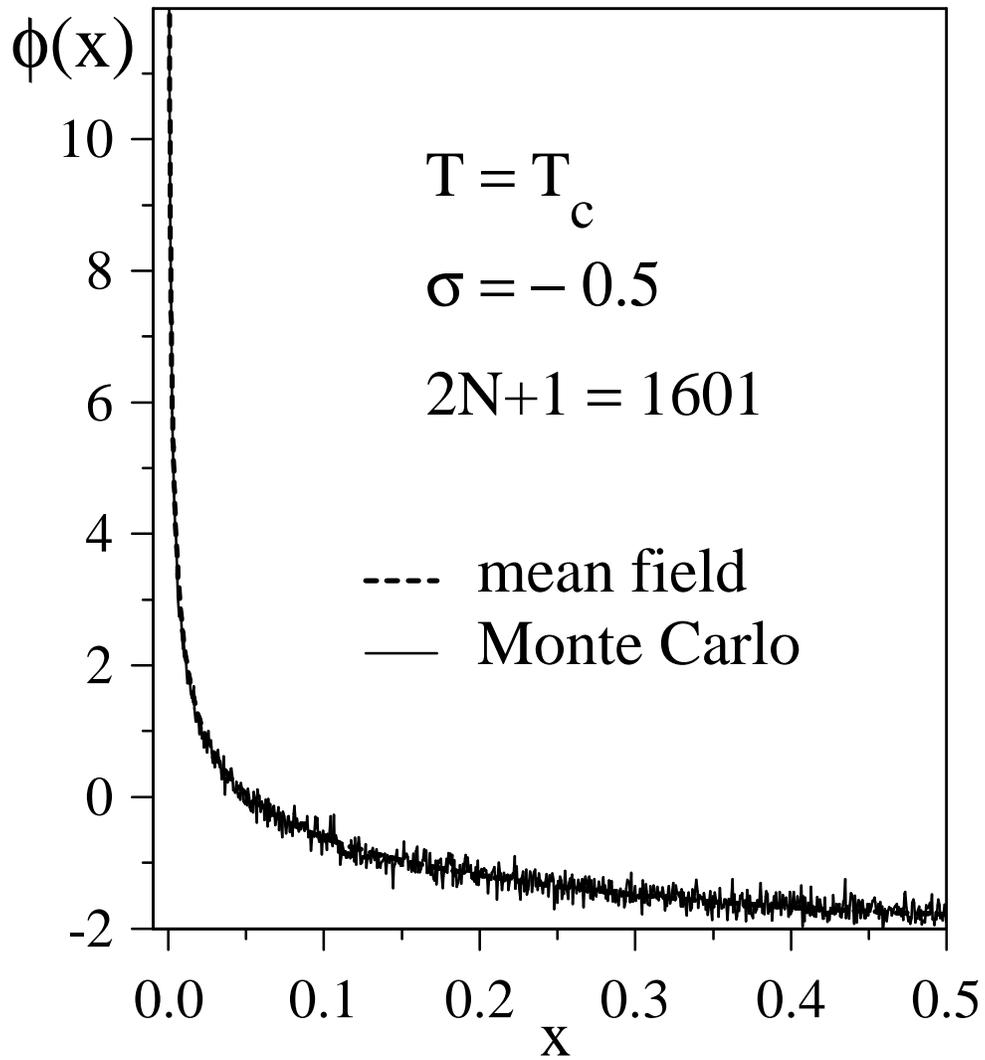
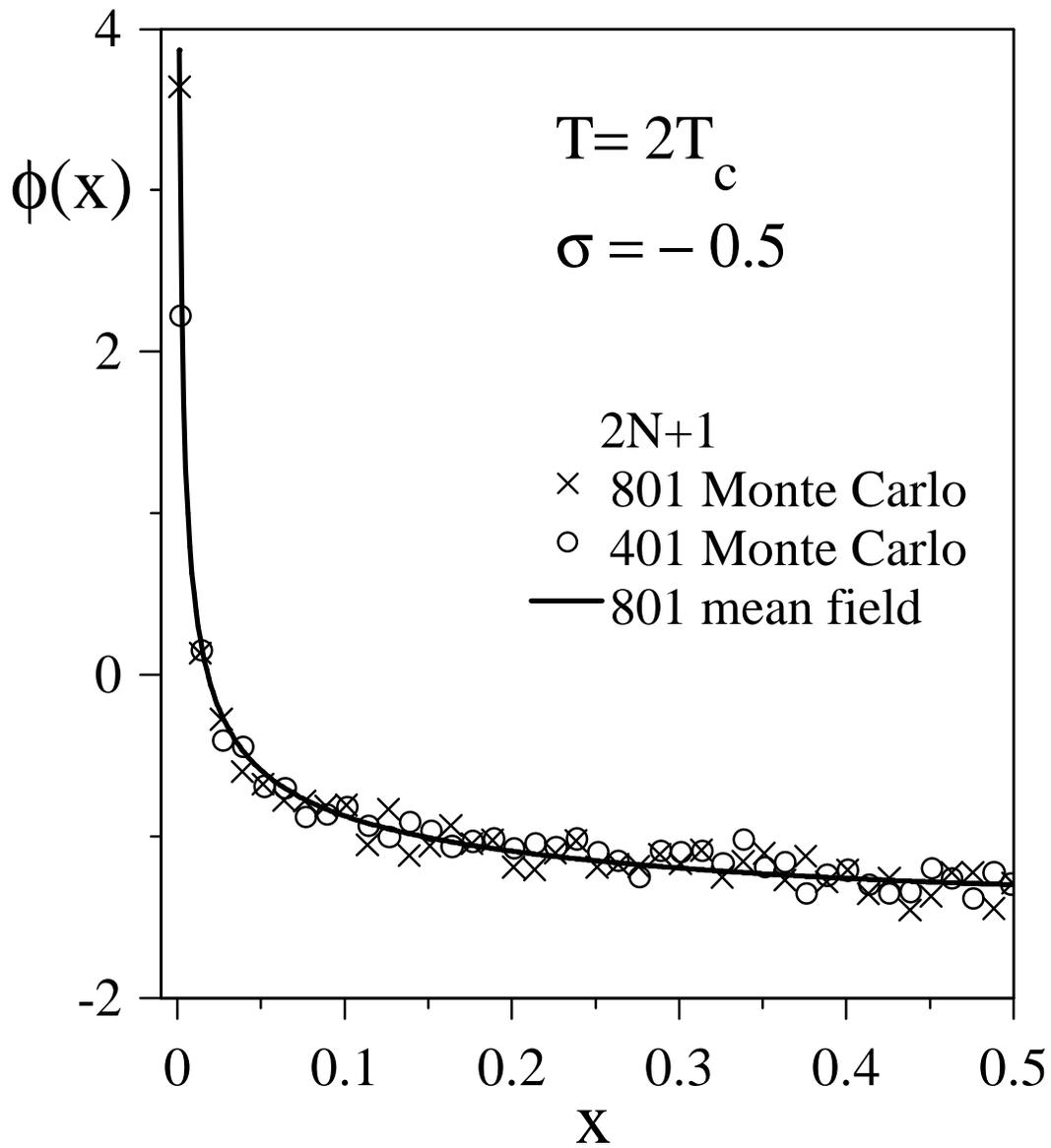
\end{document}